\documentstyle[12pt,amsfonts]{article}
\def\one{1\hskip-.37em 1}

\def\half{\textstyle{\frac{1}{2}}}

\def\H{{\cal H}}
\def\l{\lambda}
\def\p{\phi}
\def\D{{\cal D}}
\def\frak{\mathfrak}
\def\g{\gamma}
\def\s{\hskip.013cm}

\def\E{{\rm I}\hskip-.2em{\rm E}}
\def\ra{\rightarrow}
\def\tint{{\textstyle\int}}
\def\dd{\partial}
\def\o{\overline}
\def\a{\alpha}

\def\b{\begin{eqnarray*}}     
\def\e{\end{eqnarray*}}       
\def\bn{\begin{eqnarray}}     
\def\en{\end{eqnarray}}       
\def\<{\langle}
\def\>{\rangle}

\def\no{\nonumber}

\def\{{\lbrace}
\def\}{\rbrace}
\def\hg{{\hat g}}
\def\hp{{\hat\pi}}

\begin{document}

\date{}

\title{Functional Integrals in\\ Affine Quantum Gravity\footnote{Presented
at the conference ``Path Integrals - New Trends and Perspectives'',
September 2007, Dresden, Germany.}}
\author{John R.~Klauder
\footnote{Electronic mail: klauder@phys.ufl.edu}\\
Department of Physics and Department of Mathematics\\
University of Florida\\
Gainesville, FL 32611 ~~USA}


\maketitle

\begin{abstract}
A sketch of a recent approach to quantum gravity is presented which
involves several unconventional aspects. The basic ingredients
include: (1) Affine kinematical variables; (2) Affine coherent
states; (3) Projection operator approach for quantum constraints;
(4) Continuous-time regularized functional integral representation
without/with constraints; and (5) Hard core picture of
nonrenormalizability. Emphasis is given to the functional
integral expressions.

\end{abstract}




\section{Introduction}
This paper offers an introduction to the program of Affine Quantum
Gravity (AQG) and its use of functional integrals. It is important
at the outset to remark that this program is not string theory nor
is it loop quantum gravity, the two most commonly studied approaches
to quantum gravity at the present time. Although many aspects of
this approach are still to be developed, AQG seems to the author to
be more natural than most traditional views, and, moreover,
it lies closer to classical (Einstein) gravity as well. Some general
references for this paper are \cite{kla1,kla2,kla3}.

\section{Affine Kinematical Variables}
\subsubsection*{Metric positivity}
A fundamental requirement of AQG is the strict positivity of the
spatial metric. For the classical metric, this property means that
for any nonvanishing set $\{u^a\}$ of real numbers and any
nonvanishing, nonnegative test function, $f(x)\ge0$, that
 \bn  \tint f(x)\s u^a g_{ab}(x)\s u^b\s d^3\!x>0\;,  \en
where $1\le a,b \le3$.
We  also insist that this inequality holds when the classical metric field
$g_{ab}(x)$ is replaced with the
$3\times3$ operator metric field $\hg_{ab}(x)$.

\subsubsection*{Affine commutation relations}
Since the canonical commutation relations are in conflict with the
requirement of metric positivity, our initial step involves
replacing the classical ADM canonical momentum $\pi^{ab}(x)$ with
the classical mixed-index momentum $\pi^a_b(x)\equiv
\pi^{ac}(x)g_{cb}(x)$. We refer to $\pi^a_b(x)$ as the ``momentric"
tensor being a combination of the canonical {\it momen}tum and the
canonical me{\it tric}. Besides the metric being promoted to an
operator $\hg_{ab}(x)$, we also promote the classical momentric
tensor to an operator field $\hp^a_b(x)$; this pair of operators
form the basic kinematical affine operator fields, and all operators
of interest are given as functions of this fundamental pair. The
basic kinematical operators are chosen so that they satisfy the
following set of {\it affine commutation relations} (in units where
$\hbar=1$, which are normally
 used throughout): \bn
  &&\hskip.191cm[\hp^a_b(x),\,\hp^c_d(y)]=\half\s i\s[\s\delta^c_b\hp^a_d(x)-
  \delta^a_d\hp^c_b(x)\s]\,\delta(x,y)\;,\no\\
  &&\hskip.1cm[\hg_{ab}(x),\,\hp^c_d(y)]=\half\s i\s[\s\delta^c_a\hg_{bd}(x)+
  \delta^c_b\hg_{ad}(x)\s]\,\delta(x,y)\;,\label{afc}\\
&&[\hg_{ab}(x),\,\hg_{cd}(y)]=0\;.   \no \en These commutation
relations arise as the transcription into operators
of equivalent Poisson brackets for the
corresponding classical fields, namely, the spatial metric
$g_{ab}(x)$ and the momentric field
$\pi^c_d(x)\equiv \pi^{cb}(x)\s g_{bd}(x)$, along with the usual
Poisson brackets between the canonical metric field $g_{ab}(x)$ and the
canonical momentum field $\pi^{cd}(x)$.

The virtue of the affine variables and their associated commutation
relations is evident in the relation
\bn e^{ i\tint\gamma^a_b(y)\s\hp^b_a(y)\,d^3\!y}\,\hg_{cd}(x)\,
  e^{-i\tint\gamma^a_b(y)\s\hp^b_a(y)\,d^3\!y}=
\{\s e^{\g(x)/2}\s\}_c^e\,\hg_{ef}(x)\,\{\s e^{\g(x)/2}\s\}_d^f\,. \en
This algebraic relation
confirms that suitable transformations by the momentric field preserve metric
positivity.

\section{Affine Coherent States}
It is noteworthy that the algebra generated by $\hg_{ab}$ and
$\hp^a_b$ closes. These operators form
the generators of the {\it affine group} whose elements may be
defined by
  \bn U[\pi,\gamma]\equiv e^{i\tint \pi^{ab}(y)\s\hg_{ab}(y)\,d^3\!y}\,
  e^{-i\tint\gamma^a_b(y)\s\hp^b_a(y)\,d^3\!y}\;,\en
e.g., for all real, smooth $c$-number functions $\pi^{ab}$ and
$\gamma^a_b$ of compact support. Since we assume that the smeared
$\hg_{ab}$ and $\hp^a_b$ fields are self-adjoint operators, it
follows that $U[\pi,\gamma]$ are unitary operators for all $\pi$ and
$\gamma$, and moreover, these unitary operators are strongly continuous
in the label fields $\pi$ and
$\gamma$.

To define a representation of the basic operators it suffices to choose
a fiducial vector and thereby to introduce a set of affine
coherent states, i.e., coherent states formed with the help of the
affine group. We choose $|\s\eta\>$ as a normalized fiducial vector
in the original Hilbert space $\frak H$, and we consider a
set of unit vectors each of
which is given by \bn
  |\pi,\gamma\>\equiv e^{i\tint \pi^{ab}(x)\,\hg_{ab}(x)\,d^3\!x}\,
e^{-i\tint\gamma^d_c(x)
  \,{\hat\pi}^c_d(x)\,d^3\!x}\,|\s\eta\>\;.  \en
As $\pi$ and $\gamma$ range over the space of smooth
functions of compact support, such vectors form the desired set of
coherent states. The specific representation of the kinematical
operators is fixed once the vector $|\s\eta\>$ has been chosen.
As minimum requirements on $|\s\eta\>$ we impose
         \bn &&\hskip.041cm\<\eta|\hp^a_b(x)|\s\eta\>=0\;, \\
             &&\<\eta|\hg_{ab}(x)|\eta\>={\tilde g}_{ab}(x)\label{t8}\;, \en
where ${\tilde g}_{ab}(x)$ is a metric that determines the topology
of the underlying space-like surface.
As algebraic consequences of these conditions, it follows that
  \bn
   && \hskip.07cm\<\pi,\g|\s \hg_{ab}(x)\s|\pi,\g\>=\{\s e^{\g(x)/2}\s\}_a^c
   \,{\tilde g}_{cd}(x)\,\{\s e^{\g(x)/2}\s\}_b^d\equiv g_{ab}(x)
\label{t34}\;,  \\
   &&\hskip.14cm\<\pi,\g|\s{\hat\pi}^a_c(x)\s|\pi,\g\>=\pi^{ab}(x)\s g_{bc}(x)
   \equiv \pi^a_c(x)\label{t35}\;. \en
 These expectations are not gauge
invariant since they are taken in the
original Hilbert space where the constraints are not fulfilled.

By definition, the coherent states span the original, or kinematical,
Hilbert space $\frak H$, and thus we can characterize the coherent
states themselves by giving their overlap with an arbitrary coherent
state. In so doing, we choose the fiducial vector $|\s\eta\>$ so that the
overlap is given  by
 \bn &&\hskip-.3cm\<\pi'',\gamma''|\pi',\gamma'\>
   =\exp\bigg[-2\int b(x)\,d^3\!x\,\no\\
&&\hskip.1cm\times\ln\bigg(\frac{\det\{\half[g''^{ab}(x)+g'^{ab}(x)]+\half
ib(x)^{-1}
[\pi''^{ab}(x)-\pi'^{ab}(x)]\}}{\{\det[g''^{ab}(x)]\,
\det[g'^{ab}(x)]\}^{1/2}}\bigg)
\bigg] \label{t18}\en
where $b(x)$, $0<b(x)<\infty$, is a scalar density which is discussed below.

 Additionally, we
observe that $\gamma''$ and
$\gamma'$ do {\it not} appear in the explicit functional form
given in (\ref{t18}).
In particular, the smooth matrix $\gamma$ has been replaced by the
smooth matrix $g$ which is defined at every point by
 \bn  g(x)\equiv e^{\gamma(x)/2}\,{\tilde g}(x)\,
e^{\gamma(x)^T/2}\equiv\{g_{ab}(x)\}\;, \en
where $T$ denotes transpose, and the matrix ${\tilde g}(x)
\equiv \{{\tilde g}_{ab}(x)\}$ is
given by (\ref{t8}).
 The map $\gamma\ra g$ is clearly
many-to-one since $\gamma$
 has nine independent variables at each point while $g$, which is 
symmetric, has only six.
In view of this functional dependence we may denote the given
functional in (\ref{t18}) by $\<\pi'',g''|\pi',g'\>$, and henceforth
we adopt this notation. In particular, we note
that (\ref{t34}) and (\ref{t35}) become
   \bn &&
\hskip.07cm\<\pi,g|\s \hg_{ab}(x)\s|\pi,g\>\equiv g_{ab}(x)\;,  \\
   &&\hskip.143cm\<\pi,g|\s{\hat\pi}^a_c(x)\s|\pi,g\>=\pi^{ab}(x)
\s g_{bc}(x)\equiv \pi^a_c(x)\;, \en
which show that the meaning of the labels $\pi$ and $g$ is that
of {\it mean} values rather than sharp eigenvalues.

In addition, we observe that the coherent state overlap function
(\ref{t18}) is a continuous function that can 
serve as a reproducing kernel for a reproducing
kernel Hilbert space which provides a representation of the original
Hilbert space ${\frak H}$ by continuous functions of $\pi$ and $g$.
For details of such spaces, see \cite{mesh}.

\section{Projection Operator Approach for \\Quantum Constraints}
Classically, constraints are either: (i) first class, for which the
Lagrange multipliers are undetermined and must be chosen to find a
solution; or (ii) second class, for which the Lagrange multipliers
are fixed by the equations of motion.

The Dirac approach to the quantization of constraints requires
quantization before reduction. Thus the constraints are first
promoted to self-adjoint operators,
  \bn \phi_\a(p,q)\ra\Phi_\a(P,Q)\;, \en
for all $\a$, and then the physical Hilbert space ${\frak H}_{phys}$ is
defined by those vectors $|\psi\>_{phys}$  for which
   \bn \Phi_\a(P,Q)\s|\psi\>_{phys}=0\; \label{t77}\en
for all $\a$.
This procedure works for a limited set of classical first class constraint
systems, but it does not work in general and especially not for 
second class constraints.

The projection operator approach to quantum constraints involves a
slight relaxation of the Dirac procedure. Instead of insisting
that (\ref{t77}) holds exactly, we introduce a projection operator
$\E$
defined by
  \bn  \E=\E(\Sigma_\a\Phi_\a^2\le\delta(\hbar)^2)\;,  \en
where $\delta(\hbar)$ is a positive {\it regularization parameter}
and we have assumed that
$\Sigma_\a\Phi_\a^2$ is self adjoint. This relation means that $\E$
projects onto the spectral range of the self-adjoint operator
$\Sigma_\a\Phi_\a^2$ in the interval $[0,\delta(\hbar)^2]$, and then
${\frak H}_{phys}=\E\s{\frak H}$.
As a final step, the parameter $\delta(\hbar)$ is reduced as much as
required, and, in particular, when some second-class constraints are
involved, $\delta(\hbar)$ ultimately remains strictly positive. This
general procedure treats all constraints simultaneously and treats
them all on an equal basis; for details see \cite{sch}.

A few examples illustrate how the projection operator method works.
If $\Sigma_\a\s\Phi_\a^2=J_1^2+J_2^2+J_3^2$, the Casimir operator of
$su(2)$, then $0\le\delta(\hbar)^2<3\hbar^2/4$ works for this first
class example. If $\Sigma_\a\s\Phi_\a^2=P^2+Q^2$, where
$[Q,P]=i\hbar\one$, then $\hbar\le\delta(\hbar)^2<3\hbar$ covers
this second class example. If the single constraint $\Phi=Q$, an
operator whose zero lies in the continuous spectrum, then it is
convenient to take an appropriate form limit of the projection
operator as $\delta\ra0$; see \cite{sch}. The projection operator
scheme can also deal with irregular constraints such as $\Phi=Q^3$,
and even mixed examples with regular and irregular constraints such
as $\Phi=Q^3(1-Q)$, etc.; see \cite{kl-lit}.

It is also of interest that the desired projection operator
has a general, time-ordered integral representation (see \cite{kla6})
given by
  \bn  \E=\E(\!\!(\Sigma_\a\s\Phi_\a^2\le\s\delta(\hbar)^2\s)\!\!) 
=\int {\sf T}\s
  e^{-i\tint\lambda^\a(t)\s\Phi_\a\,dt}\,\D R(\lambda)\;. \label{e10}\en
The weak measure $R$ depends on the number of Lagrange multipliers,
the time interval, and the regularization parameter
$\delta(\hbar)^2$. The measure $R$ does {\it not} depend on the constraint
operators, and thus this relation is an operator identity, holding for any set
of operators $\{\Phi_\a\}$.
The time-ordered integral representation for
$\E$ given in (\ref{e10}) can be used in path-integral representations as will
become clear below.

\section{Continuous-time Regularized \\Functional Integral Representation
\\without/with Constraints} It is pedagocially useful to reexpress the
coherent-state overlap function by means of a functional integral.
This process can be aided by the fact that the expression
(\ref{t18}) is analytic in the variable $g''^{ab}(x) +i\s
b(x)^{-1}\s\pi''^{ab}(x)$ up to a factor. As a consequence, the
coherent-state overlap function satisfies a complex
polarization condition, which leads to a second-order differential
operator that annihilates it. This fact can
be used to generate a functional integral representation of the form
  \bn  &&\<\pi'',g''|\pi',g'\>
              =\exp\bigg[-2\int b(x)\,d^3\!x\,\no\\
&&\hskip1.2cm\times\ln\bigg(\frac{\det\{\half[g''^{ab}(x)+g'^{ab}(x)]+\half
ib(x)^{-1}
[\pi''^{ab}(x)-\pi'^{ab}(x)]\}}{\{\det[g''^{ab}(x)]\,
\det[g'^{ab}(x)]\}^{1/2}}\bigg)\bigg] \no\\
&&\hskip.8cm=\lim_{\nu\ra\infty}\,{\o{\cal N}}_{\nu}\,\int \exp[-i\tint g_{ab}
\s{\dot\pi}^{ab}\,d^3\!x\,dt]\no\\
  &&\hskip1.4cm\times\exp\{-(1/2\nu)\tint[b(x)^{-1}g_{ab}g_{cd}
{\dot\pi}^{bc}{\dot\pi}^{da}+
  b(x)g^{ab}g^{cd}{\dot g}_{bc}{\dot g}_{da}]\,d^3\!x\,dt\}\no\\
&&\hskip2.3cm\times[\s \Pi_{x,t}\,\Pi_{a\le
b}\,d\pi^{ab}(x,t)\,dg_{ab}(x,t)\s] \label{e20}\;.  \en 
Because
of the way the new independent variable $t$ appears in the
right-hand term of this equation, it is natural to interpret $t$,
$0\le t\le T$, $T>0$, as coordinate ``time''. The fields on the
right-hand side all depend on space and time, i.e.,
$g_{ab}=g_{ab}(x,t)$, ${\dot g}_{ab}=\dd g_{ab}(x,t)/\dd t$, etc.,
and, importantly, the integration domain of the formal measure is
strictly limited to the domain where $\{g_{ab}(x,t)\}$ is a
positive-definite matrix for all $x$ and $t$. For the boundary
conditions, we have $\pi'^{ab}(x)\equiv\pi^{ab}(x,0)$,
$g'_{ab}(x)\equiv g_{ab}(x,0)$, as well as
$\pi''^{ab}(x)\equiv\pi^{ab}(x,T)$, $g''_{ab}(x)\equiv g_{ab}(x,T)$
for all $x$. Observe that the right-hand term holds for any
$T$, $0<T<\infty$, while the left-hand and middle terms are
independent of $T$ altogether.

In like manner, we can incorporate the constraints into a functional integral
by using an appropriate form of the integral representation (\ref{e10}). 
The resultant expression has a
functional integral representation given by
 \bn  && \<\pi'',g''|\s\E\s|\pi',g'\>
      =\int \<\pi'',g''|{\bf T}\,e^{-i\tint[\s N^a\s\H_a+N\s\H\s]\,d^3\!x\,dt}
  \s|\pi',g'\>\,\D R(N^a,N)\no\\
 &&\hskip1cm=\lim_{\nu\ra\infty}{\o{\cal N}}_\nu\s\int e^{-i\tint[g_{ab}
{\dot\pi}^{ab}+N^aH_a+NH]
 \,d^3\!x\,dt}\no\\
  &&\hskip1.5cm\times\exp\{-(1/2\nu)\tint[b(x)^{-1}g_{ab}g_{cd}
{\dot\pi}^{bc}{\dot\pi}^{da}+
  b(x)g^{ab}g^{cd}{\dot g}_{bc}{\dot g}_{da}]\,d^3\!x\,dt\}\no\\
  &&\hskip2cm\times[\s \Pi_{x,t}\,\Pi_{a\le b}\,d\pi^{ab}(x,t)\,
dg_{ab}(x,t)\s   ]\,
  \D R(N^a,N)\;. \label{f39}\en
Despite the general appearance of
({\ref{f39}), we emphasize once again that this representation has
been based on the affine commutation relations and {\it not} on any
canonical commutation relations.

The expression $\<\pi'',g''|\s\E\s|\pi',g'\>$  denotes the coherent-state
matrix elements of the projection operator $\E$ which projects
onto a subspace of the original Hilbert space on which the quantum
constraints are fulfilled in a regularized fashion. Furthermore, the
expression $\<\pi'',g''|\s\E\s|\pi',g'\>$ is another continuous
 functional that can be used as a reproducing
kernel and thus used directly to generate the reproducing kernel
physical Hilbert space on which the quantum constraints are
fulfilled in a regularized manner.
Observe that  $N^a$ and $N$ denote Lagrange
multiplier fields (classically interpreted as the shift and lapse),
while $H_a$ and $H$ denote phase-space symbols (since $\hbar\ne0$)
associated with the quantum diffeomorphism and Hamiltonian
constraint field operators, respectively. Up to a surface term, 
therefore, the phase factor in the
functional integral represents the canonical action for general
relativity.

\section{Hard-core Picture of Nonrenormalizability}
Nonrenormalizable quantum field theories involve an infinite number of
distinct counterterms when approached by a regularized, renormalized
perturbation analysis. Focusing on scalar field theories, a qualitative
Euclidean functional integral formulation is given by
  \bn  S_\l(h)={\cal N}_\l\int e^{\tint h\s\p\,d^n\!x
-W_o(\p)-\l\s V(\p)}\;{\cal D}\p\;,  \en
where $W_o(\p)\ge0$ denotes the free action and $V(\p)\ge0$ the
interaction term. If $\l=0$, the support of the integral is determined by 
$W_o(\p)$;
when $\l>0$, the support is determined by $W_o(\p)+\l\s V(\p)$.
Formally, as $\l\ra0$, $S_\l(h)\ra S_0(h)$, the functional integral for 
the free theory.
However, it may happen that
   \bn  \lim_{\l\ra0} S_\l(h)=S'_0(h)\not= S_0(h)\;,  \en
where $S'_0(h)$ defines a so-called {\it pseudofree} theory.
Such behavior arises formally
if $V(\p)$ acts partially as a {\it hard core}, projecting out 
certain fields that are not
restored to the support of the free theory as $\l\ra0$ \cite{kla11}.

It is noteworthy that there exist highly idealized nonrenormalizable
model quantum field theories with exactly the behavior described;
see \cite{book}. Such examples involve counterterms not suggested 
by a renormalized perturbation analysis. It is our belief that 
these soluble models strongly
suggest that nonrenormalizable $\varphi^4_n$, $n\ge5$, models can be
understood by the same mechanism, and that they too can be properly
formulated by the incorporation of a limited number of counterterms
distinct from those suggested by a perturbation treatment. Although
technically more complicated, we see no fundamental obstacle in
dealing with quantum gravity on the basis of an analogous hard-core
interpretation.

\end{document}